\documentclass[twocolumn,english,aps,prl,superscriptaddress]{revtex4}
\usepackage{lmodern}
\usepackage[T1]{fontenc}
\usepackage[latin9]{inputenc}
\usepackage{verbatim}
\usepackage{amsmath}
\usepackage{amssymb}
\usepackage{graphicx}
\usepackage{setspace}
\usepackage{esint}
\usepackage[version=3]{mhchem}
\usepackage{graphicx,color,psfrag}

\begin{document}


\title{Dispersion engineered high-Q silicon Nitride Ring-Resonators via Atomic Layer Deposition}

\author{Johann Riemensberger}
\affiliation{\'{E}cole Polytechnique F\'{e}d\'{e}rale de Lausanne (EPFL), CH-1015 Lausanne, Switzerland}
\author{Klaus Hartinger}
\affiliation{\'{E}cole Polytechnique F\'{e}d\'{e}rale de Lausanne (EPFL), CH-1015 Lausanne, Switzerland}
\affiliation{Menlo Systems GmbH, Am Klopferspitz 19a, D-82152 Martinsried, Germany}
\author{Tobias Herr}
\affiliation{\'{E}cole Polytechnique F\'{e}d\'{e}rale de Lausanne (EPFL), CH-1015 Lausanne, Switzerland}
\author{Victor Brasch}
\affiliation{\'{E}cole Polytechnique F\'{e}d\'{e}rale de Lausanne (EPFL), CH-1015 Lausanne, Switzerland}
\author{Ronald Holzwarth}
\affiliation{Menlo Systems GmbH, Am Klopferspitz 19a, D-82152 Martinsried, Germany}
\affiliation{Max-Planck-Institut f\"ur Quantenoptik, D-85748 Garching, Germany}
\author{Tobias J. Kippenberg}
\email{tobias.kippenberg@epfl.ch}
\affiliation{\'{E}cole Polytechnique F\'{e}d\'{e}rale de Lausanne (EPFL), CH-1015 Lausanne, Switzerland}
\affiliation{Max-Planck-Institut f\"ur Quantenoptik, D-85748 Garching, Germany}

\date{\today}


\begin{abstract}
We demonstrate dispersion engineering of integrated silicon nitride based ring resonators through conformal coating with hafnium dioxide deposited on top of the structures via atomic layer deposition (ALD). Both, magnitude and bandwidth of anomalous dispersion can be significantly increased. All results are confirmed by high resolution frequency-comb-assisted-diode-laser spectroscopy and are in very good agreement with the simulated modification of the mode spectrum.
\end{abstract}

\maketitle



Silicon nitride (\ce{Si3N4}) integrated planar waveguide and ring resonator structures \cite{Henry1987} are attractive platforms for resonant nonlinear frequency conversion \cite{Levy2010}. Moreover \ce{Si3N4} has been used for ultra-low loss integrated waveguides \cite{Bauters2011b}, particularly in the optical telecom band, where well established silicon on insulator waveguides \cite{Foster2006, Koos2007} suffer from two-photon and free carrier absorption. Besides low absorption, the waveguide dispersion plays a central role for parametric frequency conversion. In particular this applies to microresonator based optical frequency comb generation via the $\chi^{(3)}$ non-linearity ("Kerr-combs") [6], which has recently also been demonstrated in \ce{Si3N4} [7]. In this scheme a set of equidistant optical frequencies is generated with a spacing corresponding to the free spectral range (FSR) of the resonator. Such integrated and CMOS-compatible microresonator frequency combs potentially offer a level of compactness and integration that is presently not attainable in mode-locked laser based combs and can moreover access GHz repetition rates. Microresonator based frequency combs are promising for applications like the calibration of astronomical spectrographs \cite{Steinmetz2008}, on chip optical interconnects \cite{Levy2010} and optical arbitrary waveform generation \cite{Ferdous2011}. It is well known that the attainable spectral bandwidth of Kerr-combs is limited by the resonator dispersion, that leads to a wavelength dependent FSR of the resonator \cite{Del'Haye2011, Okawachi2011}. The dispersion may be described as $ D_{2} = \omega_{l-1}-2\omega_{l}+\omega_{l+1}$, where $\omega_l$ is the frequency of the fundamental resonance with azimuthal mode number $l$. The dispersion $D_2$ in microresonators is related to the group velocity dispersion of the structures via $ \beta_{2} = - D_{2}/\left(2 \pi R D_{1}^3 \right) $, with $ D_{1}= \omega_{l+1}-\omega_{l} $ being the FSR of the resonator and $R$ the ring radius. It has been shown that the phase noise characteristics of Kerr-comb generators is related to the ratio of cavity-decay-rate and dispersion \cite{Herr2011} and a sufficiently large anomalous dispersion is advantageous for low phase noise operation of Kerr-combs (intrinsically low phase noise combs are generated for $\sqrt(\kappa/D_2 \sim 1$). Dispersion in integrated ring resonators is composed of material dispersion (which is normal for \ce{Si3N4} at all visible and near-infrared wavelengths), geometric dispersion due to the waveguide cross-section \cite{Turner2006}, as well as an additional contribution from the finite resonator radius as observed in whispering-gallery-mode microcavities \cite{Del'Haye2009}. In previous numerical studies it has been shown that silicon ($n = 3.5$) waveguide dispersion can be decreased and flattened through a conformal coating with \ce{Si3N4}  ($n = 1.99$) \cite{Liu2008}. In this letter we demonstrate experimentally dispersion engineering, that is a targeted increase of magnitude and bandwidth of anomalous dispersion in integrated \ce{Si3N4} ring resonators through conformal coating of hafnium dioxide (\ce{HfO2} $n = 2.03$). This coating has the advantage of very small scattering at the \ce{HfO2}/\ce{Si3N4} interface. Furthermore scattering at the \ce{HfO2}/\ce{SiO2} interface could potentially be reduced because the surface roughness of \ce{Si3N4} is masked by the \ce{HfO2} coating \cite{Alasaarela2011}.



\begin{figure}[hbp]
\includegraphics[width=\linewidth -2cm]{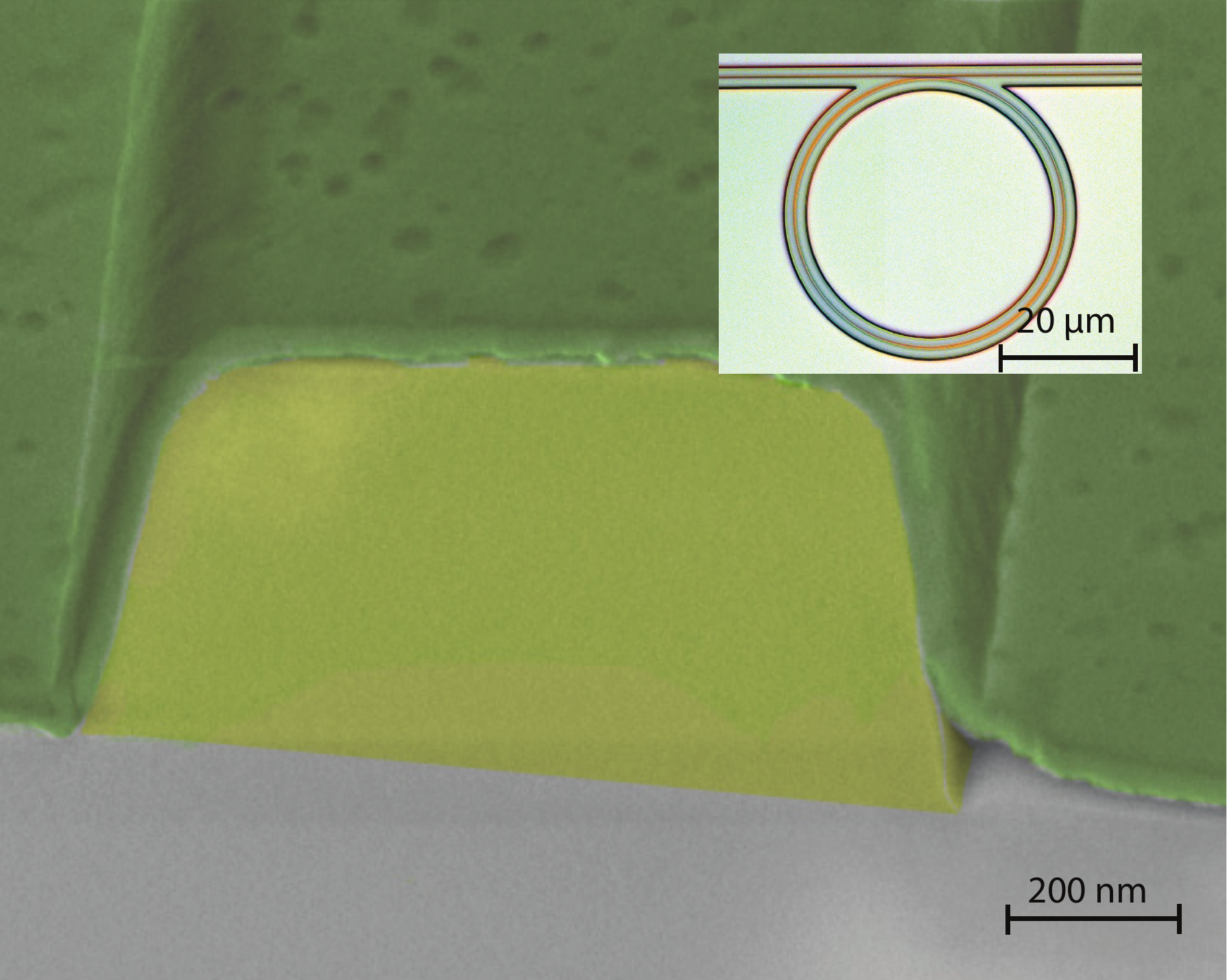}
\caption{Scanning electron micrograph of the cross-section of an \ce{HfO2} (green) coated \ce{Si3N4} (yellow) resonator. The detachment of the \ce{HfO2} film on the lower right side of the waveguide was caused by the cleaving process. Inset: Optical Micrograph of a 50 $\mu$m radius ring resonator and coupling waveguide.}
\label{pics}
\end{figure}

The waveguides and resonators are fabricated as follows. First, a bottom cladding layer of 4 $\mu$m thick silicon dioxide (\ce{SiO2}) is grown onto a standard silicon wafer via wet oxidation. A 750 nm thick stoichiometric \ce{Si3N4} layer is grown in two steps with low-pressure-chemical-vapor-deposition (LPCVD) with intermediate cooldown to room temperature \cite{Gondarenko2009} to reduce tensile stress in the layer. After electron beam lithography (using ZEP520A resist), the waveguide and resonator patterns are transferred into the waveguide core layer via reactive-ion-etching (RIE) in \ce{SF6} and \ce{CH4} RF plasma. The etch parameters have been carefully optimized to achieve low roughness while maintaining verticality and selectivity. The cross-section of the ring resonators after etching is 750 nm high and 1700 nm wide at the base with a sidewall angle of approximately 78$^\circ$. The resist is then stripped in an \ce{O2} Plasma. Excess \ce{Si3N4} (not forming waveguides) is removed to reduce intrinsic stress via additional etching steps. After cleaning using the Piranha process the waveguides and resonators are thermally annealed in a nitrogen atmosphere at a temperature of 1200 $^o$C to eliminate residual hydrogen and to reduce optical absorption in the near-infrared \cite{Henry1987}. \ce{HfO2} with a thickness of 55 nm is grown on top of the waveguides and resonators by ALD from tetrakis(ethylmethylamino)hafnium (TEMAH) and water (c.f. Fig. \ref{pics}). 
Finally, 3~$\mu$m thick low-temperature oxide is deposited on top on the structures. The critically coupled optical quality factors Q of the resonators are up to 1 million with \ce{HfO2} clad resonators having slightly smaller quality factors (up to $Q\approx 6\cdot 10^5$) due to the formation of small irregularities on the surface (c.f. Fig. \ref{pics}a,d) during deposition. These are expected to vanish with further optimized ALD parameters and wafer preparation. The circumference of the chips is defined via an additional photolithography step and coupling facets are etched with RIE. To facilitate cleaving of the wafer and clear space to approach the coupling facet with tapered-lensed-fibers for input and output coupling of light, a 130 $\mu$m deep anisotropic silicon RIE etch is performed. To reduce losses from fiber-to-chip coupling the bus waveguides are tapered \cite{Almeida2003} to a width of 150 nm and terminated 6 $\mu$m away from the etched coupling facet.



\begin{figure}[hbp]
\includegraphics[width=\linewidth]{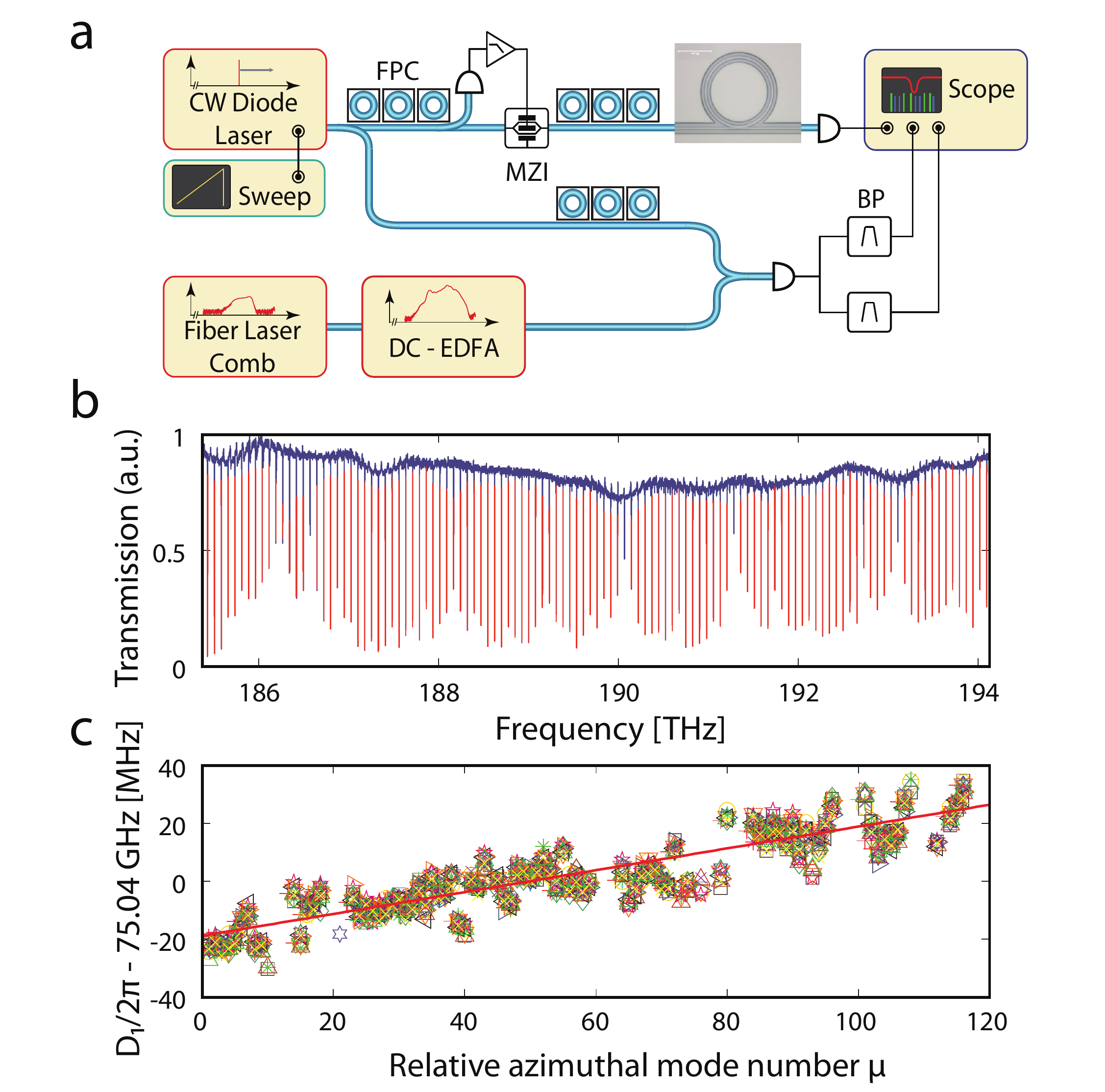}
\caption{a) Setup for frequency comb assisted diode laser spectroscopy; FPC, fiber polarization controller; MZI, Mach-Zehnder intensity modulator; BP, bandpass; DC-EDFA, dispersion compensated erbium-doped-fiber-amplifier. b) Calibrated transmission spectrum of a 55 nm ALD-coated ring resonator with radius 301 $\mu$m and input polarization aligned to the TM11 mode c) Wavelength dependence of the free-spectral-range (FSR) over the measurement span of 8.7 THz, corresponding to 120 subsequent FSRs, based on 16 consecutive scans (symbols) in alternating direction. The red line is a linear fit from which both FSR $D_{1}/2\pi$ of 75.04 GHz and dispersion $D_{2}/2\pi$ of 350 kHz are determined.}
\label{experiment}
\end{figure}

\begin{figure}[hbp]
\includegraphics[width=\linewidth]{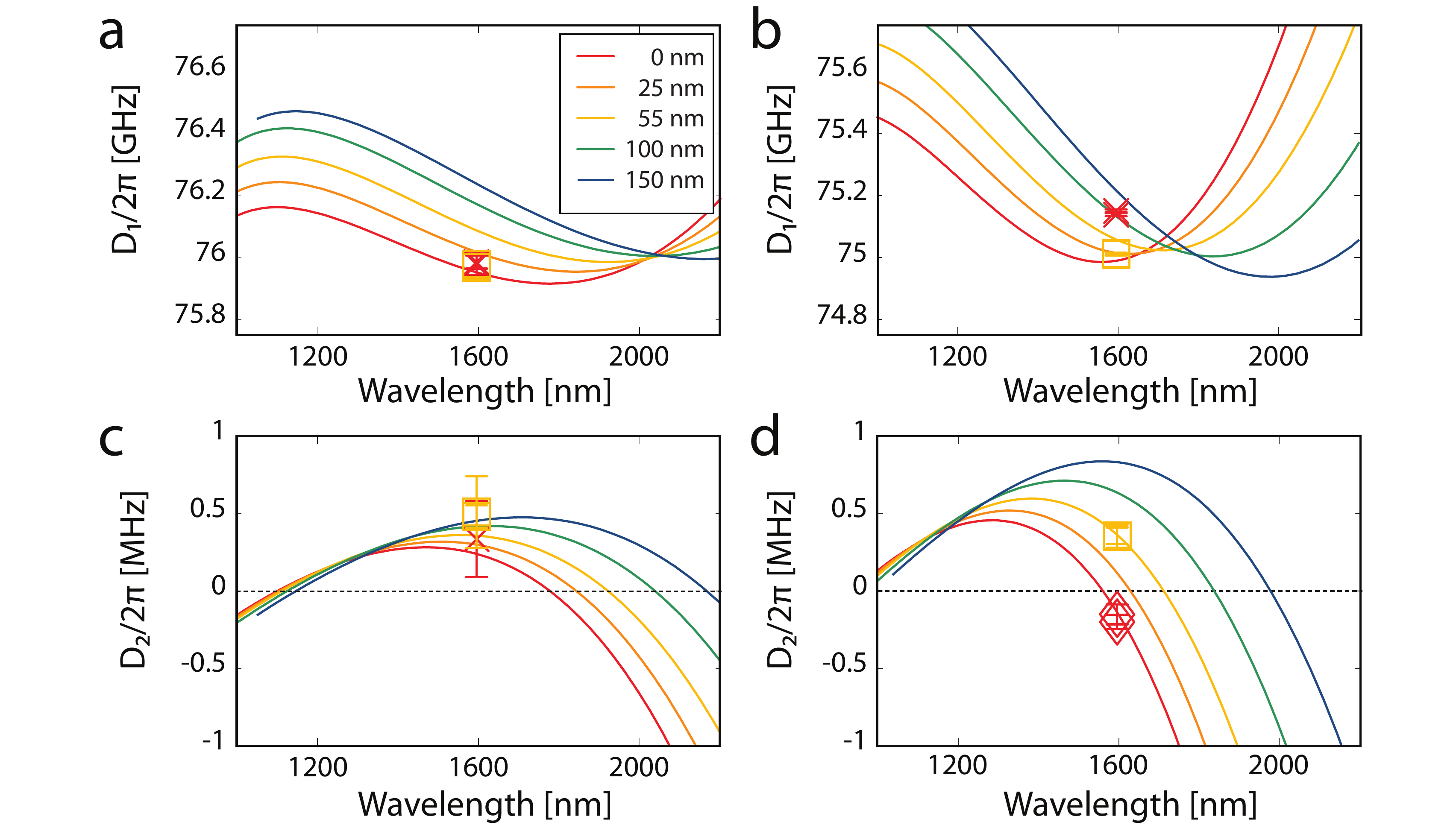}
\caption{a) Simulations (lines) and measurements (symbols) of FSR $D_{1}/2\pi$ and dispersion $D_{2}/2\pi$ of fundamental TE (a,c) and TM modes (b,d) for different \ce{HfO2} ALD coating thickness applied to 301 $\mu$m ring resonators.}
\label{aldthickness}
\end{figure}

\begin{figure}[hbp]
\includegraphics[width=\linewidth]{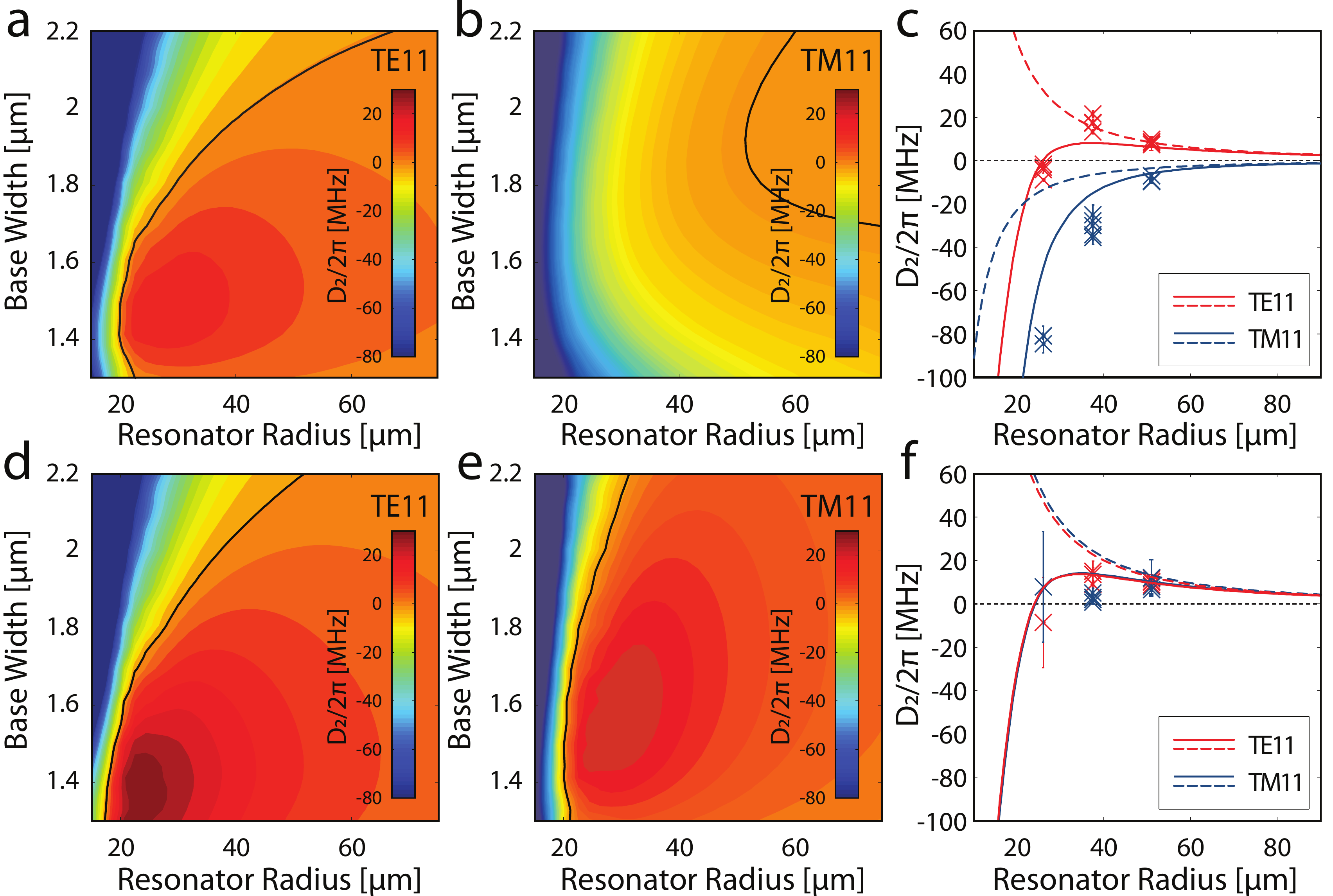}
\caption{Simulated resonator dispersion $D_2/2\pi$ at 1.55 $\mu$m for the TE11(a) and TM11(b) mode as function of ring radius and ring base width. The black solid line marks the transition from normal to anomalous dispersion. c) Comparison of dispersion simulations (lines) and measurements (symbols) for TE11 and TM11 mode for 24 $\mu$m, 37 $\mu$m and 50 $\mu$m ring resonators with a base width of 1700 nm. Dashed lines show the simulated dispersion neglecting geometric dispersion of a bent geometry. d,e,f) Same as (a,b,c) but with 55 nm \ce{HfO2} coating.}
\label{resrad}
\end{figure}

The resonator dispersion is characterized employing frequency-comb-assisted-diode-laser spectroscopy \cite{Del'Haye2009} (c.f. Fig. \ref{experiment}a).
Here, a mode-hop-free external cavity diode laser (ECDL) is scanned between 1545 nm and 1610 nm. The spectrum of the resonator in transmission is recorded while a beat note with a self-referenced, stabilized fiber-laser-frequency-comb with repetition rate of 250 MHz is generated. The beat signal is split and bandpass filtered at 30 MHz and 75 MHz to create two sets of accurate frequency calibration markers. All traces are recorded on a fast oscilloscope in peak detection mode with 10 million sampling points. To extend the span of the measurement the frequency comb is amplified and broadened in SMF28. The intensity of the scanning laser is actively stabilized using a Mach-Zehnder interferometric intensity modulator. The measurement span is limited by the mode-hop free scanning range of the ECDL and the signal-to-noise ratio of the beat note generation. The resonance positions are fitted with symmetric double Lorentzian functions to account for modal coupling of counter-propagating modes \cite{Kippenberg2002}. The cross-section of the ring resonators are designed to feature anomalous dispersion at 1.55 $\mu$m and support both TE(M)11 and TE(M)21 modes. While in larger resonators (e.g. 301 micron radius) the excitation of higher transverse order modes is suppressed by the phase matching condition, these modes can be excited in smaller resonator geometries. Different mode families are identified via their FSR (c.f. Fig. \ref{aldthickness}a, b). These different FSR lead to periodic crossings of mode families \cite{Del'Haye2009} and to the formation of hybrid modes due to modal coupling \cite{Carmon2008}. These effects are observable as spectral variation of the resonance depth (c.f. Figure \ref{experiment}b). Data points affected by modal crossings are discarded before the dispersion $D_{2}$ is extracted from a linear fit of the measured FSR over the full spectral span (c.f. Fig. \ref{experiment}c). Another systematic error arises from wavelength dependent waveguide transmission due to interference of higher order and fundamental waveguide modes and asymmetric Fano resonance line shapes \cite{Chiba2005}. Each dispersion measurement corresponds to 16 scans in alternating scan direction to cancel thermal shifts of resonance positions \cite{Carmon2004} and reduce the statistical error from calibration and double-Lorentzian fitting. Different measurement points in Fig. \ref{aldthickness} and Fig. \ref{resrad} correspond to different resonators with different coupling strength.



We simulate the dispersion of integrated \ce{Si3N4} based ring resonators with a fully-vectorial axially-symmetric finite-element model \cite{Oxborrow2007} employing a commercial FEM solver (COMSOL Multiphysics). The model includes the angled sidewalls as well as the shape of the conformal coating on top of the waveguide edges (c.f. Fig. \ref{pics}). The model also accounts for the bent resonator waveguide geometry affecting the geometric dispersion component. The material dispersion is determined via spectral ellipsometry and included iteratively (similar to Ref. \cite{Del'Haye2009}) into the simulations. The results of the simulations for varied \ce{HfO2} coating thickness are shown in Figure \ref{aldthickness}. We find, that, both the magnitude and bandwidth of anomalous dispersion can be significantly increased (c.f. Fig. \ref{aldthickness}c,d).

  The simulations are compared to measurements of 301 $\mu$m ring resonators as shown in Figure \ref{experiment}c. We measure an increase of anomalous dispersion $D_2/2\pi$ at 1.55 $\mu$m from 340 kHz to 490 kHz for the TE11 mode for a coating thickness of 55 nm. The TM11 mode, i.e. the mode with the major electric field component out of the resonator plane, is more strongly affected by the coating. We measure that uncoated ring resonators have normal dispersion (-180 kHz) while  resonators coated with 55 nm \ce{HfO2} feature anomalous dispersion (350 kHz). While comparable results could also be achieved employing a thicker \ce{Si3N4} core layer, which is difficult because of the large intrinsic stress of stochiometric \ce{Si3N4}, or by partial or complete removal of the \ce{SiO2} cladding, which would sacrifice full encapsulation of the device, coating with \ce{HfO2} facilitates large anomalous dispersion without aforementioned drawbacks. For a constant group-velocity-dispersion $\beta_{2}$ it is expected that the total dispersion scales as $D_{2}/2\pi \propto 1/R^2$. However, for whispering gallery mode resonators it has been observed that smaller wavelengths travel closer to the outer rim of the resonator and additional normal dispersion is introduced \cite{Del'Haye2009}. Figure \ref{resrad} presents measurements and simulations, which show that this effect also applies for ring resonators with strong radial confinement, such as integrated waveguide rings. Similar effects have been numerically studied for bent silicon waveguides \cite{Zhang2010}.  In Figure \ref{resrad} simulations of the dispersion $D_2/2\pi$ as function of waveguide width and ring radius, which are the two lithographically determined geometry parameters for given \ce{Si3N4} layer thickness, are presented.

 We find a maximum of anomalous dispersion as function of ring radius and ring base width. Coating with \ce{HfO2} leads to a increased maximum which is shifted towards narrower ring base widths and smaller ring radii. The shift towards narrower waveguide width is due to the transition of a waveguide like mode to a disc like mode. The shift towards smaller radii is due to the increased anomalous waveguide dispersion. The predictions of the numerical simulations are confirmed by measurements on ring resonators with a base width of 1700 nm and ring radii of 24 $\mu$m, 37 $\mu$m and 50 $\mu$m (c.f. Fig. \ref{resrad}c,f). 



In summary, we have demonstrated both numerically and experimentally that dispersion of silicon nitride based waveguides can be engineered to be more anomalous through conformal coating with \ce{HfO2} which is in particular favourable for applications such as low phase noise Kerr-comb generation. To facilitate measurements, we have extended the measurement span of frequency comb assisted diode laser spectroscopy to more than 8 THz with MHz resolution and accuracy to resolve the dispersion of integrated ring resonators (values as small as x100 below the cavity decay rate can be measured). Using this unique method of directly measuring the dispersion of travelling wave microresonators with large FSR we have been able to demonstrate that the total \ce{Si3N4}-ring resonator dispersion for small ring radii is always normal. Our results also show how \ce{HfO2} coating can be used to extend the difficult fabrication of thick \ce{Si3N4} based waveguide devices effectively reducing the necessary thickness of high stress, low loss stochiometric \ce{Si3N4} layers.

\section*{Acknowledgements}

This work was supported by the NanoTera program of the SNF under the acronym micro-Comb, by DARPA under the QuASAR program and by the Swiss National Science Foundation (SNF). K. H. acknowledges support under the Marie Curie IAPP Fellowship. The samples used in this work were fabricated at the Center for Microfabrication (CMi) at EPFL.


\end{document}